\documentclass[a4paper,10pt,fleqn]{article}
\usepackage[dvips,colorlinks,bookmarksopen,bookmarksnumbered,linkcolor=black,citecolor=black,urlcolor=black,breaklinks]{hyperref}
\usepackage{graphicx}
\usepackage{breakurl}
\usepackage[authoryear]{natbib}
\usepackage{times}
\usepackage{mathrsfs}
\usepackage{color}
\begin{document}

\title{Seamless phase II/III clinical trials using early outcomes for treatment or subgroup selection: Methods and aspects of their implementation}

\author{Tim Friede$^1$, Nigel Stallard$^2$ and Nicholas Parsons$^2$ \\ \\
\normalsize $^1$Department of Medical Statistics, University Medical Center G\"ottingen,\\ 
\normalsize Humboldtallee 32, 37073 G\"ottingen, Germany\\
\normalsize $^2$Division of Health Sciences, Warwick Medical School,\\ 
\normalsize The University of Warwick, Coventry, UK}

\date{12 January 2019}

\maketitle

\begin{abstract}
Adaptive seamless designs combine confirmatory testing, a domain of phase III trials, with features such as treatment or subgroup selection, typically associated with phase II trials. They promise to increase the efficiency of development programmes of new drugs, e.g. in terms of sample size and / or development time. It is well acknowledged that adaptive designs are more involved from a logistical perspective and require more upfront planning, often in form of extensive simulation studies, than conventional approaches. Here we present a framework for adaptive treatment and subgroup selection using the same notation, which links the somewhat disparate literature on treatment selection on one side and on subgroup selection on the other. Furthermore, we introduce a flexible and yet efficient simulation model that serves both designs. As primary endpoints often take a long time to observe, interim analyses are frequently informed by early outcomes. Therefore, all methods presented accommodate both, interim analyses informed by the primary outcome or an early outcome. The R package {\tt asd}, previously developed to simulate designs with treatment selection, was extended to include subpopulation selection (so-called adaptive enrichment designs). Here we describe the functionality of the R package {\tt asd} and use it to present some worked-up examples motivated by clinical trials in chronic obstructive pulmonary disease and oncology. The examples illustrate various features of the R package providing at the same time insights into the operating characteristics of adaptive seamless studies. 
\end{abstract}






\section{Introduction}
\noindent 
There is a long history of application of sequential methods in clinical trials that allow the monitoring of accumulating data at a series of interim analyses \citep{JennisonTurnbull99, Whitehead97}.  Whilst most early work focussed on the aim of stopping the trial as soon as sufficient evidence has been obtained, this body of work has rapidly expanded to include the use of interim data for other design adaptations, including sample size reestimation \citep{FriedeKieser06, Proschan09}.  Over the past years, there has been considerable interest in using interim analyses for selection of treatments, with less effective treatments dropped from the study, or for selection of patient subgroups, with recruitment following an interim analysis limited to subgroup(s) in which a promising effect is indicated \citep{Bauer16, Pallmann18}.

A major statistical challenge in the development of such methods is the control of the type I error rate when adaptations are made on the basis of data that will also be included in the final analysis.  For treatment selection, most methods proposed fall into two groups.  The first builds on the work of Thall et al. (1988, 1989) using the group-sequential method, but require that a single treatment continues along with a control beyond the first stage \citep{StallardTodd03} or that the number of treatments at each stage is specified in advance \citep{StallardFriede08}.  The second group of methods is based on the combination test approach of \citet{bk94}.  Such methods are more flexible (see, for example, \citet{BauerKieser99}, \citet{Posch05}, and \citet{Bretz06}), but may be less powerful in some settings \citep{FriedeStallard08}.  Magirr et al. (2012) proposed a group-sequential method that does allow completely flexible treatment selection, though this may be at the cost of conservatism and an associated loss in power. \citet{Koenig08} showed how the conditional error principle of \citet{MuellerSchaefer01} may be used to extend the Dunnett test \citep{Dunnett55} to a two-stage design with flexible treatment selection.  This has been shown to compare well in terms of power with competing methods \citep{FriedeStallard08}.  

There is a smaller body of work on clinical trials with subgroup selection (see \citet{Brannath09, Jenkins11, Wassmer15}). This has mainly used the combination testing approach, though methods based on the conditional error principle \citep{Friede12, Stallard14, Placzek18} and the group-sequential approach \citep{Magnusson13} have also been proposed. An overview is provided in \citet{Ondra16}.

Earlier work had assumed that the adaptations would be informed by the primary outcome, which is also used for hypothesis testing. From a practical perspective, however, this can be a strong limitation. In particular in chronic diseases clinically meaningful endpoints might take some time to observe, which means that most or all patients are recruited by the time the primary outcome is observed for the first patients. This is illustrated for example by \citet{Chataway11} in the context of secondary progressive multiple sclerosis. As a consequence, adaptations need to be based on early outcomes for adaptive desings to be feasible in these situations. Therefore, some adaptive seamless designs have been extended to allow the use of short-term endpoint data for decision making at interim \citep{Stallard10, Friede11, Jenkins11, Friede12, Kunz14, Kunz15, Stallard15}.

It is acknowledged that these more complex designs require intensive simulation studies in the planning to evaluate their operating characteristics \citep{Benda10, Friede10}. A limitation to the use of adaptive methods in practical applications is often the availability of software to enable construction and evaluation of appropriate study designs and to conduct the final analysis. A number of commercial software packages including ADDPLAN and EAST are available for this purpose. Although some R packages for group-sequential designs including including {\tt gsDesign} and {\tt gscounts} and adaptive group-sequential designs such as {\tt rpact} are available from CRAN, there is still a shortage of comprehensive freely available software for adaptive seamless designs with treatment or subgroup selection, with the exception of the R package {\tt{asd}} developed to plan clinical trials with treatment selection \citep{Parsons12}.

The aim of this paper is to present the approaches for treatment and subgroup selection in a unified notation. By expressing the subgroup selection problem in a similar setting to that of treatment selection, the R package {\tt{asd}}, originally developed for treatment selection, can be used for designs of both types. The methods implemented are based on the combination testing approach.  Therefore, the designs obtained are fully flexible, controlling the type I error rate for any data-driven adaptation (treatment / subgroup selection as well as sample size adaptation). 

Although the employed test procedures do not make use of the joint distribution of the tests statistics, we derive the joint distribution to develop a fairly general and efficient simulation model. This is based on standardized test statistics only requiring the assumption of multivariate normal distributions for the test statistics, but not the individual observations. Furthermore, early outcomes informing the interim decisions are incorporated, since this is often important in practice as explained above. The application of the methods using the R package {\tt asd} will be illustrated by clinical trials in chronic obstructive pulmonary disease (COPD) and oncology with treatment and subgroup selection, respectively.

\section{Methods}

\subsection{Notation and hypotheses}
We consider first the setting of treatment selection designs.  The study is conducted in up to $J \ge 2$ stages.  In the first stage, patients are randomised between $K$ experimental treatments and a control treatment. In a general setting, suppose that observations of the primary outcome from group $k$, $k=0, \ldots, K$, where $k=0$ corresponds to the control group, have a distribution depending on some parameter $\mu_k$, and that it is desired to test the family of hypotheses $H_k: \theta_k=0$, $k=1, \ldots, K$, where $\theta_k = \mu_k - \mu_0$.  Let $p_{kj}$ denote a p-value for the test of hypothesis $H_k$ based on the data from patients first observed in stage $j$. These data may not be available at the time of the interim analysis following stage $j$, but become available only later on in the trial \citep{Friede11}. As we will explain in Section \ref{sec_selrules} below, the interim decisions need not necessarily be based on the primary outcome but could, in principle, make use of any available outcome while still maintaining control of the type I error rate. 

Next we consider the setting of subgroup selection designs. Suppose that a single experimental treatment is to be compared with a control but that patients can be categorised as belonging to one or more predefined subgroups. Interest is focused on the treatment effect in the full population and each subgroup, in particular in the subgroup with the largest treatment effect.  

As part of the aim of this paper is to make explicit the similarities between methodology for treatment and subgroup selection designs, we will use similar notation for both settings when this does not cause confusion.  Thus suppose that observations come from patients in $K$ subgroups labeled $k, k=1, \ldots, K$.  Suppose that data for patients in subgroup $k$ receiving treatment $r, r=0,1$ have a distribution depending on some parameter $\mu_{kr}$.  Setting $\theta_k = \mu_{k1} - \mu_{k0}$, it is desired to test the family of hypotheses $H_k: \theta_k=0$, $k=1, \ldots, K$.  As above, $p_{kj}$ denotes a p-value for the test of hypothesis $H_k$ based on the data from patients with an outcome first observed in stage $j$.

\subsection{Interim selection rules} \label{sec_selrules}
The testing strategies described here control the type I error rate for any selection rule, but it is good practice to specify selection rules in advance to enable calculation of operating characteristics including sample size and power. In the following we introduce some possible examples of interim selection rules based on the final outcome, but understand that these could equally be applied to an early outcome as we will explain below.

One obvious way to proceed would be to select the treatment or subpopulation which performs best in terms of some statistic $Z_{j,k}$ (where $Z_{j,k}$ is some estimator of $\theta_k$ following stage $j$). However, this might not be wise in situations where sample sizes are relatively small given the differences between the treatments, since there is a rather high risk of picking some other but the optimal treatment or subpopulation. The so-called $\epsilon$-rule proposed by \citet{Kelly05} selects all treatments or subpopulations with statistics $Z_{j,k}$ for which $Z_{j,k} \ge \max_i Z_{j,i} - \epsilon$ (assuming that larger values of $Z_{j,k}$ are better). For $\epsilon=0$ this rule reduces to selecting the maximum only. For large $\epsilon$ no selection takes place as all treatments or subgroups are carried forward. Otherwise varying numbers of treatments or subpopulations are carried forward into the next stage. 

Multi-arm studies including several doses of an experimental drug motivate another selection rule. In some indications it is not uncommon to select not one but two doses for confirmatory testing in phase III. The COPD study discussed in more detail in Section \ref{sec_examples} is a good example for this. A more generalized version of this rule would be to select the best $K^\star$ out of $K$ treatments or subpopulations where $K^\star$ would be specified in advance.

The selection of treatments or subgroups in interim analyses could be informed by the primary outcome or, if this is not feasible, by an early outcome. The early outcome must not necessarily fulfill all requirements of a surrogate endpoint \citep{Burzykowski05} as weaker conditions might suffice. \citet{Chataway11} coined the phrase of a ``biologically plausible'' outcome that ``gives some indication as to whether the mechanism of action of a test treatment is working as anticipated''. Nevertheless, for the operating characteristics of the adaptive seamless design the correlation between the early and the final outcomes on an individual patient level as well as the treatment effects on both the early and the final outcome (population level) are relevant as we will see below.

\subsection{Error rate control via the closed testing procedure} \label{sec_closedtest}
In either the treatment selection or the subgroup selection setting, the problem has been posed in such a way that it is desired to test the null hypotheses $H_k: \theta_k \le 0, k=1, \ldots, K$.  It is desirable to conduct these hypothesis tests such as to control the familywise error rate in the strong sense, that is control of the probability of rejection of any true hypothesis within this family, at some specified level, $\alpha$.  Strong error rate control my be achieved through a closed testing procedure in which, denoting by $H_\mathscr{K}$ the intersection hypothesis $\cap_{k \in \mathscr{K}} H_k$, all hypotheses $H_\mathscr{K}$ for $\mathscr{K} \subseteq \{1, \ldots, k\}$ are tested at nominal level $\alpha$, and $H_k$ rejected if and only if $H_\mathscr{K}$ is rejected at this level for all $\mathscr{K} \ni k$ \citep{Marcus76}.

Application of the closed testing procedure requires a test of the intersection hypothesis $H_\mathscr{K}$ for each $\mathscr{K} \subseteq \{1, \ldots, k\}$.  These hypothesis tests must also combine evidence from the different stages in the trial.  This may be achieved through the use of a combination testing method, as described in detail in the next subsection.

\subsection{The combination testing method and early stopping} \label{sec_combtest}
Extending the notation introduced above, let $p_{\mathscr{K} j}$ denote a p-value for a test of the hypothesis $H_\mathscr{K}$ based on data from patients with an outcome first observed at stage $j$.  By construction, under $H_\mathscr{K}$, $p_{\mathscr{K} j} \sim U[0,1]$, or if a conservative test is used, $p_{\mathscr{K} j}$ is stochastically no smaller than a $U[0,1]$, for all $j$ and $\mathscr{K}$.  We also assume that the conditional distribution of $p_{\mathscr{K} j}$ given $p_{\mathscr{K} 1}, \ldots, p_{\mathscr{K} j-1}$ is stochastically no smaller than a uniform for all $p_{\mathscr{K} 1}, \ldots, p_{\mathscr{K} j-1}$ for all $j$, which is also referred to as the \textit{p-clud} condition \citep{Brannath02}. The condition is satisfied if the p-values from different stages are independent.  

The p-values from the different stages can be combined using a number of combination functions \citep{bk94, LehmacherWassmer99} to give test statistics $C_{\mathscr{K} j}(p_{\mathscr{K} 1}, \ldots, p_{\mathscr{K} j})$ which, given the assumptions above regarding the distributions of the p-values, have known distributions under $H_\mathscr{K}$ irrespective of adaptations made to the study design.  These test statistics may thus be used as the basis of a test of hypothesis $H_\mathscr{K}$ \citep{BauerKieser99, Bretz06}.

Although a number of combination functions have been proposed, we will use the inverse normal combination function \citep{LehmacherWassmer99}, which is equivalent to the method of \citet{Cui99}.  This gives test statistics $C_{\mathscr{K} j} = \sum_{j^\prime=1}^j w_{j^\prime} \Phi^{-1} (1-p_{\mathscr{K} j^\prime})$ where $w_1, \ldots, w_J$ are specified in advance.  Given the distributional assumptions under $H_\mathscr{K}$, $C_{\mathscr{K} j}$ are distributed as, or are stochastically no larger than, a multivariate normal distribution with mean zero, $var(C_{\mathscr{K} j}) = w^2_1 + \cdots + w_j^2$ and $cov(C_{\mathscr{K} j},C_{\mathscr{K} j^\prime}) = w_1^2 + \cdots + w_{\min\{j,{j^\prime}\}}^2$.

Following \citet{Posch05} we assume that hypotheses once they are dropped cannot be rejected anymore, which results in conservative tests. Furthermore, applying the closed testing principle outlined in Section \ref{sec_closedtest} in an adaptive design the p-value $p_{\mathscr{K} j}$ is replaced by $p_{\mathscr{K} \cap \mathscr{I}_j j}$ where $\mathscr{I}_j$ is the set of hypotheses carried forward into stage $j$ \citep{Posch05}.  

If interim analyses are used for adaptation of the design but not for stopping the trial, a final test of $H_\mathscr{K}$ may be based on $C_{\mathscr{K} J}$.  More generally, a sequential test of $H_\mathscr{K}$ may be conducted based on the joint distribution of $C_{\mathscr{K} 1}, \ldots, C_{\mathscr{K} J}$, rejecting $H_\mathscr{K}$ if $C_{\mathscr{K} j} \ge u_{\mathscr{K} j}$ for some critical values $u_\mathscr{K} = (u_{\mathscr{K}1}, \ldots, u_{\mathscr{K}J})^T$.  To simplify the notation, $u_\mathscr{K}$ and $u_{\mathscr{K}j}$ will generally be written as $u$ and $u_j$ when it is clear which hypothesis is being tested.  The single constraint that the overall error rate should be at most $\alpha$ is insufficient to determine $u$ uniquely.  A common approach in sequential analysis is to specify the type I error to be spent at each interim analysis, and to find $u_{\mathscr{K} 1}, \ldots, u_{\mathscr{K} J}$ to satisfy  
\begin{equation}
\label{eq:spending}
pr_{H_\mathscr{K}}(C_{\mathscr{K} j^\prime} \ge u_{\mathscr{K} j^\prime}, \mbox{ some } j^\prime \le j) = \alpha^*_j
\end{equation}
where $\alpha^*_1 \le \cdots \le \alpha^*_J = \alpha$ are either specified in advance \citep{Slud82} or depend on the observed information in some predetermined way \citep{Lan83}.  Critical values $u_{\mathscr{K} 1}, \ldots, u_{\mathscr{K} J}$ satisfying (\ref{eq:spending}) can be found recursively with $u_{\mathscr{K} j}$ found directly from the joint distribution of $C_{H_\mathscr{K} 1}, \ldots, C_{H_\mathscr{K} J}$ via a numerical search once $u_{\mathscr{K} 1}, \ldots, u_{\mathscr{K} j-1}$ are known.  Computational details are given by e.g. \citet{JennisonTurnbull99}.

Construction of the test statistics $C_{\mathscr{K} 1}, \ldots, C_{\mathscr{K} J}$, on which the sequential test is based, requires specification of p-values $p_{\mathscr{K} 1}, \ldots, p_{\mathscr{K} J}$ for testing $H_\mathscr{K}$.  For elementary hypotheses, that is when $| \mathscr{K} | = 1$, $p_{\mathscr{K} j}$ can be obtained from a standard test, such as a t-test for normally distributed data or a chi-squared test for binary data.  When $| \mathscr{K} | > 1$, $p_{\mathscr{K} j}$ should be calculated so as to allow for the multiple comparisons implicit in testing $H_\mathscr{K}$.  In the treatment selection setting, as comparisons are with a common control, a Dunnett test \citep{Dunnett55} may be used. In the subgroup selection setting, the simple Bonferroni procedure, Simes' procedure \citep{Brannath09} or the Spiessens-Debois test \citep{SpiessensDebois10}, a Dunnett-type test with a generalized covariance structure, may be used. In each case the level of adjustment depends on the size of $\mathscr{K}$.

\section{Simulation model}
As eluded to in the Introduction, simulations are often required in the planning phase of a study to chose certain design options such as sample sizes or interim selection rules. Here we propose a simulation model that is efficient in the sense that population statistics are generated rather than individual patient data. Therefore, computation times do not increase with larger sample sizes. 

In a wide variety of settings it is possible to obtain statistics that are, at least asymptotically, normally distributed with known variance. When a series of interim analyses is conducted, these statistics have a multivariate normal distribution, with statistics obtained at different interim analyses correlated because of the use of earlier data in later stages. In the setting of treatment or subgroup selection, these multivariate normal distributions may be extended to give the joint distribution of statistics corresponding to different treatment comparisons or for treatment effects in different subgroups.  To be clear, these statistics are not actually necessarily the test statistics used to test the hypotheses but are rather some estimator of $\theta_k$. Also the distributional forms assumed in the simulations need not be used as the basis of hypothesis tests, since these, as described in Section \ref{sec_combtest}, can use the combination testing approach.

Since, as indicated above, we will also be considering adaptations to the trial design based on the observation of short-term endpoint data, it is helpful to further extend the simulation models to include test statistics calculated based on different endpoints. The distributions of the resulting test statistics are described briefly in this section. Additional detail is given in the appendix.

\subsection{Treatment selection}
We consider first treatment selection designs.  Let $\hat \theta_{kj}$ denote the estimate of the treatment effect, $\theta_k$, for treatment $k$ based on the data available at the $j$th interim analysis, and $I_{kj}^{-1}$ denote the variance of this estimate.  As these estimates are often, at least asymptotically normally distributed, our model will be based on an assumption of this distributional form.  Assume that $I_{kj}$ does not depend on $k$, and so may be denoted by $I_j$, and that the correlation between different treatment comparisons with a common control group is $1/(1+\lambda)$, as will often be the case if we have $1:\lambda$ randomisation.  Setting $S_{kj} = \hat \theta_{kj} I_{kj}$, we then have   
\[
(S_{11}, \ldots, S_{K1}, \ldots, S_{1J}, \ldots, S_{KJ})^\prime \sim
N \left ( I \otimes \theta, \Sigma^{(GS)}(I) \otimes \Sigma^{(CS)}_K(\frac{1}{1+\lambda}) \right )
\]
where $I$ and $\theta$ denote respectively the vectors $(I_1, \ldots, I_J)^\prime$, and $(\theta_1, \ldots, \theta_K)^\prime$ and 
\[
\Sigma^{(CS)}_K(r) = 
\left (
\begin{array}{cccc}
1 & r & \cdots & r \\ 
r & 1 & \ddots & \vdots \\ 
\vdots & \ddots & \ddots & r\\
r & \cdots & r & 1 \\ 
\end{array}
\right )
\]
and 
\[
\Sigma^{(GS)}(I)^{(1)} = 
\left (
\begin{array}{cccc}
I^{(1)}_1 & I^{(1)}_1 & \cdots & I^{(1)}_1 \\ 
I^{(1)}_1 & I^{(1)}_2 & \cdots & I^{(1)}_2 \\
\vdots & \vdots & \ddots & \vdots \\
I^{(1)}_1 & I^{(1)}_2 & \cdots & I^{(1)}_J\\
\end{array}
\right )
\]
 denote variance matrices of the $K \times K$ complex symmetric form and of that obtained in the usual group-sequential setting.

\subsection{Subgroup selection}
In the subgroup selection setting, let $\hat \theta_{kj}$ denote the estimate of the treatment effect, $\theta_k$, for subgroup $k$ based on the data available at the $j$th interim analysis.  Again, assume that these estimates are normally distributed, let $I_{kj}^{-1}$ denote the variance of this estimate and assume that $I_{kj}$ does not depend on $k$, and so may be denoted by $I_j$. In the simplest case, in which $K=2$ and subgroup 2 is the whole population and subgroup 1 a proportion of size $\tau$, setting $S_{kj} = \hat \theta_{kj} I_{kj}$, we have   
\[
(S_{11}, S_{21}, \ldots, S_{1J}, S_{2J})^\prime \sim
N \left ( I \otimes \theta, \Sigma^{(GS)}(I) \otimes \Sigma^{(GS)}\left( \begin{array}{c} \tau \\1 \end{array}\right) \right ).
\]

Extensions to the case when a short-term endpoint is also considered, and to more complex subgroup selection settings are discussed in the appendix.

\section{Software implementation in R}

\subsection{R package {\tt{asd}}}
The simulation models for two-stage treatment selection and subgroup selection designs, described in Section 3, can be implemented in the R \citep{R} package {\tt{asd}} \citep{Parsons12}, which is available from the Comprehensive R Archive Network (CRAN) at {\tt http://cran.r-project} {\tt .org/package=asd}). This package comprises a number of functions that allow the properties of seamless phase II/III clinical trial designs, potentially using early outcomes, for treatment or subgroup selection to be explored and evaluated prior to a study commencing. 
An earlier version of this package, without the extension to subgroup designs, enhanced options for a wider range of outcome measures and more complete output model description was described previously by \citet{Parsons12}. The general structure of the code comprises a set of base functions that implement lower level tasks such as hypothesis testing, treatment selection and closed testing procedures. The base functions, which have been described previously in the setting of treatment selection designs \citep{Parsons12}, have been modified in the latest version of the \texttt{asd} package to work with both subgroup and treatment selection designs. The higher level user facing function \texttt{asd.sim} has been replaced by two new functions \texttt{treatsel.sim} and \texttt{subpop.sim} that are called directly by the user to implement hypothesis testing and simulations. The more general functions, \texttt{gtreatsel.sim} and \texttt{gsubpop.sim} can in principle be called directly by the user, although due to the more complex input structure this is generally not recommended.

\subsection{Function {\tt{subpop.sim}}}
\citet{Friede12} extended the previously described CT approaches, for co-primary analyses in a single pre-defined subgroup and the full population, using the methods proposed by \citet{SpiessensDebois10} to control the FWER in the subgroup and the full population, and also proposed a novel method to obtain a critical value for the definitive test using a CEF approach; full details of these methods are given by \citet{Friede12}. The function {\tt{subpop.sim}} implements all the methods for subgroup selection in adaptive clinical trials reported by \citet{Friede12} and subsequent correspondence \citep{Friede12corr}. The authors described and explored the performance of a number of methods in the setting described here, distinguishing between two distinct approaches to control the familywise error rate (FWER), a combination test (CT) method \citep{Brannath09, Jenkins11} and a conditional error function method (CEF). 

\subsubsection{Input arguments}
An overview and brief description of the input arguments available for {\tt{subpop.sim}} is shown in Table \ref{tab:functdescrip}. The combination test (CT) methodology described by \citet{Friede12} can be implemented in {\tt{subpop.sim}} using either the \citet{SpiessensDebois10} (SD), Simes or Bonferroni testing procedures to control the FWER. These and the CEF approach can be selected using the {\tt{method}} argument to {\tt{subpop.sim}} using the following options; (i) CT-SD ({\tt{method="CT-SD"}}), (ii) CT-Simes ({\tt{method="CT-Simes"}}), (iii) CT-Bonferroni ({\tt{method="CT-Bonferroni"}}) and (iv) CEF ({\tt method} {\tt ="CEF"}).

The syntax providing the group sample sizes, for stages 1 and 2, for a putative trial design and effect sizes for early and final outcomes is consistent across the available outcome types and comprises a list of the selected options. For instance, for a design where the required sample size per treatment arm is 100 for stage 1 and 200 for stage 2 would be implemented using the following expression, {\tt n=list(stage1=} {\tt 100,stage2=200)}. The assumption, in the current version of {\tt{asd}}, is that the sample size in the control arm of the study is the same as in the treatment arm. The default setting is that if the subgroup only is selected at the interim analysis at stage 1, then the sample size in this group is the same as it would be if the study continued with the full population. If an increase in the sample size in the subgroup were planned if this group only were selected (enrichment) then this can be implemented by adding an additional item to the list to indicate this; for instance, if we wanted a sample size of 200 in the subgroup in this setting, irrespective of the subgroup prevalence, then we would modify the previous argument to the following {\tt{n=list(stage1=100,enrich=200,stage2=200)}}. 

Effect sizes for early and final outcomes are also given as a list using expressions of the following structure, where the first element of each vector is the effect size in the subgroup and the second element is the effect size in the full population, {\tt{effect=list(early=c(0.3,0.1),final=c(0.3,0.1))}}. So here we are assuming an effect size of 0.3 in the subgroup and 0.1 in the full population; the effect size in the control group is set by default to be zero. The default setting is for normal outcomes for both early and final outcomes, {\tt{outcome=list(early="N",final="N")}}, and the effects are interpreted given these options. The allowed options for outcome types are normal ({\tt{N}}), time-to-event ({\tt{T}}) and binary ({\tt{B}}), and all combinations of these are allowed for early and final outcome measures. Generally, it is assumed that higher means ({\tt{N}}) and lower event rates ({\tt{B}} or {\tt{T}})  are better. A detailed description of these options is left for the following section describing function {\tt{treatsel.sim}}; the options described there are analogous to those available for {\tt{subpop.sim}}. In the simpler setting where group selection is based purely on the final outcome is required, this can be implemented by setting the effect sizes for the early and final outcomes to be equal and setting the correlation between the early and final outcomes to one (i.e. {\tt{corr=1}}). 

The subgroup prevalence is set by a single argument ({\tt{sprev}}), where for instance {\tt{sprev=0.5}} indicates that the subgroup comprises half of the full population. The function {\tt{subpop.sim}} randomly generates test statistics (with a seed number set using {\tt{seed}}) and accumulates results from usually a large number of simulations that must be set using the {\tt{nsim}} option (default setting, {\tt{nsim=1000}}). The prevalence can be either fixed at the set value ({\tt{sprev.fixed=TRUE}}) or allowed to vary ({\tt{sprev.fixed=FALSE}}) using a single realization of the binomial random variate generation function {\tt{rbinom}}, at the set values for the sample size and subgroup prevalence, at each simulation.

Subgroup selection at interim is implemented using the so-called threshold selection rule \citep{FriedeStallard08, Friede12} ({\tt{select="thresh"}}), that is such that if the difference between the test statistics for the subgroup and the full population $\Delta\leq l_1$ then the subgroup only is tested at the end of stage 2, if $\Delta> l_2$ the full population only is tested and otherwise both subgroup and full populations are tested. The threshold rule limits $(l_1,l_2)$ are set using the argument {\tt{selim}} which is a vector of standard deviation multiples; if for instance large limits are set (e.g. {\tt{selim=c(-10,10)}}) then both subgroup and full populations will always be tested at the trial endpoint, whereas if {\tt{selim=c(0,0)}} only the test regarding the population with the largest test statistic at interim is taken into stage 2. Intermediate values for {\tt{selim}} between these extremes provide more flexible selection options. The weight for the CT approaches, if unset, is given by $n_{stage 1}/(n_{stage 1}+n_{stage 2})$ and the test level is set by default to 0.025 ({\tt{level=0.025}}).

\begin{table}[htb]
\begin{center}
\caption{Brief description and available input arguments to R functions {\tt{subpop.sim}} and {\tt{treatsel.sim}}}\label{tab:functdescrip} 
\centering
\begin{tabular}{l p{12cm} l  l}
\hline
$Argument$ & $Description$\\
\hline
{\tt{n}}&List giving sample sizes for each treatment group at stage 1 (interim) and stage 2 (final) analyses; {\tt{subpop.sim}} has an additional optional list item for enrichment\\
{\tt{effect}}&List giving effect sizes for early and final outcomes; an optional {\tt{control}} argument can also be included for {\tt{subpop.sim}}\\
{\tt{outcome}}&List giving outcome type for early and final outcomes; options for normal ({\tt{N}}), time-to-event ({\tt{T}}) and binary ({\tt{B}}) are currently available\\
{\tt{nsim}}&Number of simulations ({\tt{nsim}}$<1\times10^{7}$)\\
{\tt{sprev}}&Subgroup prevalence (${0}<${\tt{sprev}}$<{1}$); for function {\tt{subpop.sim}} only\\
{\tt{sprev.fixed}}&Subgroup prevalence can be either fixed or allow to vary at each simulation ({\tt{TRUE}} or {\tt{FALSE}}); for function {\tt{subpop.sim}} only\\
{\tt{corr}}&Correlation between early and final outcomes (${-1}\leq${\tt{corr}}$\leq{1}$)\\
{\tt{seed}}&Seed number for simulations\\
{\tt{select}}&Method for treatment selection; for {\tt{subpop.sim}} a threshold rule with limits using argument {\tt{selim}} and for {\tt{treatsel.sim}} one of seven available options (see text) with additional arguments {\tt{epsilon}} and {\tt{thresh}} as necessary\\
{\tt{ptest}}&A vector of treatments for specific counts of the number of rejections; for function {\tt{treatsel.sim}} only\\
{\tt{method}}&Methodology used for simulations; for {\tt{treatsel.sim}} either {\tt{invnorm}} or {\tt{fisher}} and for {\tt{subpop.sim}} either
{\tt{CT-SD}}, {\tt{CT-Simes}}, {\tt{CT-Bonferroni}} or {\tt{CEF}} (see text)\\
{\tt{fu}}&Follow-up options ({\tt{TRUE}} or {\tt{FALSE}}); for function {\tt{treatsel.sim}} only\\
{\tt{weight}}&User defined stage 1 weight (${0}\leq${\tt{weight}}$\leq{1}$)\\
{\tt{level}}&Test level for simulations (${0}\leq${\tt{level}}$\leq{1}$)\\
{\tt{file}}&File name to dump output; if unset will default to R console\\
\hline
\end{tabular}
\end{center}
\end{table}

\subsubsection{Output}
The output to {\tt{subpop.sim}} first gives a summary of the simulation model, including expected values for the test statistics at each stage of the study. The main summary table reports the number of times that hypotheses $H_0^{\{S\}}$,  $H_0^{\{F\}}$,  $H_0^{\{S,F\}}$ were selected for testing and rejected when the subgroup ($S$), the full population ($F$) or both were tested. Output from {\tt{subpop.sim}} is by default directed to the usual R console, but to save more detailed summaries of the simulation model, output can be directed to a file using this as an argument to the {\tt{file}} function (e.g. {\tt{file="output.txt"}}). Section \ref{sec_examples} shows how {\tt{subpop.sim}} is used in a practical setting with example data.

\subsection{Function {\tt{treatsel.sim}}}

Function {\tt{treatsel.sim}} replaces and generalizes the previous function {\tt{asd.sim}}, with the name change made to make it much more explicit that the code implements only simulations for treatment selection designs for multi-arm studies. Much of the syntax and model set-up is consistent between {\tt{treatsel.sim}} and {\tt{subpop.sim}}. 

\subsubsection{Input arguments}
An overview and brief description of the input arguments available for {\tt{treatsel.sim}} is shown in Table \ref{tab:functdescrip}. One aspect of the design set-up that differs considerably between {\tt{treatsel.sim}} and {\tt{subpop.sim}} is the coding of the treatment effects. Treatment effect sizes are given for the control group $\theta_{0}$ and the test treatment or treatments $\theta_{k}$ and as vectors for both early and final outcomes; for instance {\tt{effect=list( early=c(0,0.1,0.2,0.1), final=c(0,0.1,0.2,0.3))}} indicates that there are three test treatments with effect sizes 0.1, 0.2 and 0.1 for the early outcome and 0.1, 0.2 and 0.3 for the final outcome respectively; the control is set to 0 for both outcomes. There is no limit to the number of test treatment groups. However, in practice our experience is that the code runs slowly for designs with eight or more treatment groups. The setting of the effect sizes can be clarified further by considering the available options for the outcome types (normal {\tt{N}}, time to event {\tt{T}}, and binary {\tt{B}}), set using the option {\tt{outcome}}, with the default being to have both early and final outcomes normal ({\tt{outcome=list(early="N",final="N")}}), with all nine combinations available. Generally, it is assumed that higher means ({\tt{N}}) and lower event rates ({\tt{B}} or {\tt{T}}) are better. 

For normal outcomes the test statistics for the simulation model for the $K$ test treatments, relative to the control group, are given by $\sqrt{n/2}\times(\theta_{k}-\theta_{0})$. For time-to-event outcomes, effects are interpreted as minus log hazard rates, with the control $\theta_{0}$ set to zero and test statistics are given by $\sqrt{o_{k}/4}\times(\theta_{k}-\theta_{0})$, where the expected total number of events in the control and treatment groups $o_{k}$ for treatment $k$ is calculated under an assumed exponential model to be $n\times{(1-\exp(-\exp(-\theta_{0})))}+n\times{(1-\exp(-\exp(-\theta_{k})))}$. Binary effects are characterized by log odds ratios, $\sqrt{1/o_{k}+1/(n-o_{k})+1/o_{0}+1/(n-o_{0})}\times(\theta_{k}-\theta_{0})$, where $\theta_{k}$ is minus the log odds of the event and the observed number of events in treatment group $k$ is $o_{k}=n\times 1/(1+\exp(\theta_{k}))$. Some care must be taken when setting-up the simulation model, as clearly the interpretation of the {\tt{effect}} argument to {\tt{treatsel.sim}} is dependent on the options selected for {\tt{outcome}}.

The {\tt{method}} argument to {\tt{treatsel.sim}} allows either the inverse normal ({\tt{invnorm}}) or Fisher's ({\tt{fisher}}) combination test and the logical follow-up argument ({\tt{fu}}) determines whether (i) patients in the dropped treatment groups are removed from the trial and unknown test statistics in the dropped treatments are set to $-\infty$ at stage 2 ({\tt{fu=FALSE}}) or (ii) patients are kept in the trial and followed-up to the final outcome, in the same manner as the patients recruited in stage 1 in the selected treatment groups ({\tt{fu=TRUE}});  \citet{Friede11} called option (i), the default setting, discontinued follow-up and option (ii) complete follow-up.
Seven treatment selection rules based on stage 1 test statistics are available in {\tt{treatsel.sim}}, and are chosen with the {\tt{select}} argument; (i) select all treatments ({\tt{select=0}}), (ii) select the maximum ({\tt{select=1}}), (iii) select the maximum two ({\tt{select=2}}), (iv) select the maximum three ({\tt{select=3}}), (v) flexible treatment selection using the $\epsilon$-rule \citep{FriedeStallard08, Friede11}, with additional argument ({\tt{epsilon}}) ({\tt{select=4}}), (vi) randomly select a single treatment ({\tt{select=5}}) or (vii) select all treatments greater than a threshold, with the additional argument ({\tt{thresh}}) ({\tt{select=6}}).

The only additional argument available for {\tt{treatsel.sim}}, that has not been covered in the section describing {\tt{subpop.sim}}, is {\tt{ptest}}. This is a vector of valid treatment numbers for determining specific counts for the number of simulations that reject the null hypothesis; for instance, for three test treatments and {\tt{ptest=c(1,3)}}, {\tt{treatsel.sim}} will count and report the number of rejections of one or both hypotheses for testing treatments 1 and 3 against the control, in addition to the number of rejections of each of the elementary hypotheses.

\subsubsection{Output}
The output from {\tt{treatsel.sim}} first gives a summary of the simulation model, including expected values for the test statistics at each stage of the study. The main summary tables reports (i) the number of treatments selected at stage 1, (ii) treatment selection at stage 1, that is how often each treatment was selected, (iii) counts of hypotheses rejected at study endpoint, for each of the elementary hypotheses ($H_0^{\{1\}}, H_0^{\{2\}},\ldots,H_0^{\{K\}}$) and (iv) the number of times that one or more than one of the hypotheses identified in {\tt{ptest}} is rejected. Section \ref{sec_examples} shows how {\tt{treatsel.sim}} is used in a practical setting with example data.

\section{Examples} \label{sec_examples}
In this section we illustrate the methods described above by two example studies using the R package {\tt asd}. The first is a multi-arm randomised controlled trial in COPD with treatment selection;  which will be considered in Section \ref{sec_copd}. As a second example we consider trials in oncology with time-to-event outcomes and subgroup selection; this will be considered in Section \ref{sec_onco}.

\subsection{Clinical trial in COPD with treatment selection} \label{sec_copd}
\citet{Barnes11} and \citet{Donohue10} report a seamless adaptive design with dose selection, which was also discussed elsewhere \citep{Cuffe14}. Patients were randomized to four doses of indacaterol ($75 \mu g$, $150 \mu g$, $300 \mu g$ and $600 \mu g$), active controls and placebo control. For the purpose of illustration, we ignore the active controls in the following and use the observed results to illustrate the design process. The primary outcome was the percentage of days of poor control over 26 weeks. As recruitment for the entire study took only 6 months, the interim treatment selection could not be based on the primary endpoint. Trough forced expiratory volume in 1 second (FEV1) at 15 days was identified as a suitable early outcome to inform the interim analysis. From Figure 1 of \citet{Barnes11}, difference in trough FEV1 compared to placebo at 15 days is $150ml$, $180 ml$, $210ml$ and $200ml$ for the indacaterol doses $75 \mu g$, $150 \mu g$, $300 \mu g$ and $600 \mu g$, respectively. Reported 95\% confidence intervals suggest 2 standard errors of treatment difference is approximately $60ml$, so assuming equal stage 1 sample sizes n1=110, then the standard deviation of the measurements is approximately $220ml$. Standardized effect sizes are thus approximately 0.68, 0.82, 0.95 and 0.91 for the indacaterol doses $75 \mu g$, $150 \mu g$, $300 \mu g$ and $600 \mu g$, respectively. Regarding the final outcome days of poor control (\%) over 26 weeks, we gather from the report of stage 2 results at http://clinicaltrials.gov/show/NCT00463567 that the placebo rate was 35\%. Let us assume rates of 31\%, 30\%, 28\% and 29\% for the four doses of indacaterol ($75 \mu g$, $150 \mu g$, $300 \mu g$ and $600 \mu g$). Based on reported standard errors, we estimate the standard deviation to be approximately 30\%. Hence, the standardized effect sizes are approximately 0.13 ($75 \mu g$), 0.17 ($150 \mu g$), 0.23 ($300 \mu g$) and 0.20 ($600 \mu g$). The approximate sample sizes per arm were 100 patients in stage 1 and 300 patients in stage 2. Since the aim was to select two doses of indacaterol at the interim analysis to take into stage 2, we consider an overall sample size for the two stage of $5 \times 100 + 3 \times 300 = 1400$ patients. Furthermore, a moderate positive correlation between early and final outcomes of 0.4 is assumed.

In the following we will consider three settings: (i) continuous early and final outcomes and selecting always two doses for the second stage, as in the original study; (ii) continuous early and final outcomes, as in the original study, but for the purpose of illustration we vary the selection rule using the threshold rule to allow varying numbers of doses being taken forward for confirmatory testing in the second stage; (iii) as in particular final outcomes are non-normal and early and final outcomes are on different scales we assume in another setting that the final is binary and not continuous.

\subsubsection{Continuous early and final outcomes: Selecting always two doses for the second stage}
As an example we consider a fixed total sample size of 1400 patients, which we can allocate between the two stages, with either more or less resources for each stage. We consider the following combinations of $n_1=10, 25, \dots, 175$ and $n_2=450, 425, \dots, 175$ with $5 \times n_1 + 3 \times n_2 = 1400$. Each one of these sample size options can be tested using the following implementation of the {\tt treatsel.sim} function for the setting with 100 patients per arm in stage 1 and 300 patients per arm in stage 2:

{
\begin{verbatim}
treatsel.sim(n=list(stage1=100,stage2=300),
      effect=list(early=c(0,0.68,0.82,0.95,0.91),
      final=c(0,0.13,0.17,0.23,0.20)),
      outcome=list(early="N",final="N"),
      nsim=10000,corr=0.4,seed=145514,select=2,
      level=0.025,ptest=c(3,4))
\end{verbatim} }

This code sets the sample sizes for each stage, and provides the effect estimates as described above, for a normal early outcome (``{\tt N}'') and a normal (``{\tt N}'') final outcome. The correlation is set to 0.4, and the number of simulations to 10,000. The select=2 options implements the rule that chooses the two treatments with the largest test statistics an interim. The test level is set to 0.025 and the {\tt ptest} options allows us to count rejections for either or both of the doses $300 \mu g$ and $600 \mu g$. Results from running this code are as follows (omitting the parts describing the setup):

{\footnotesize \begin{verbatim}
simulation of test statistics: 
expectation early = 4.8 5.8 6.7 6.4 
expectation final stage 1 = 0.9 1.2 1.6 1.4 and stage 2 = 1.6 2.1 2.8 2.4 
weights: stage 1 = 0.5 and stage 2 = 0.87 

number of treatments selected at stage 1: 
              n               % 
     1        0            0.00 
     2    10000          100.00 
     3        0            0.00 
     4        0            0.00 
 Total    10000          100.00 

treatment selection at stage 1: 
              n               % 
     1      383            3.83 
     2     3282           32.82 
     3     8661           86.61 
     4     7674           76.74 

hypothesis rejection at study endpoint: 
              n               % 
    H1      183            1.83 
    H2     2067           20.67 
    H3     7206           72.06 
    H4     5541           55.41 

reject H3 and/or H4 = 8469 :  84.69%
\end{verbatim} }

The first part of the output provides a summary of the model set-up, and the second part values of the test statistics used in the simulations. The squared weights are calculated, in this case, as 100/400 and 300/400. The results are summarized in the three lower tables. The first indicates that two treatments were always selected at stage 1, the second gives the number of simulations that each treatment was selected and the third gives the number of simulations that the elementary hypotheses were rejected. The final statement gives the number of simulations that at least one of the treatments picked using the {\tt ptest} options were rejected. Assigning the output for this function to an object that we for the sake of illustration call simply output, then the summaries described here can be accessed directly, for instance for plotting data or other analysis, using the syntax {\tt output\$count.total}, {\tt output\$select.total}, {\tt output\$reject.total} and {\tt output\$sim.reject}.

\begin{figure}[ht]
\centering
\includegraphics[width=12cm]{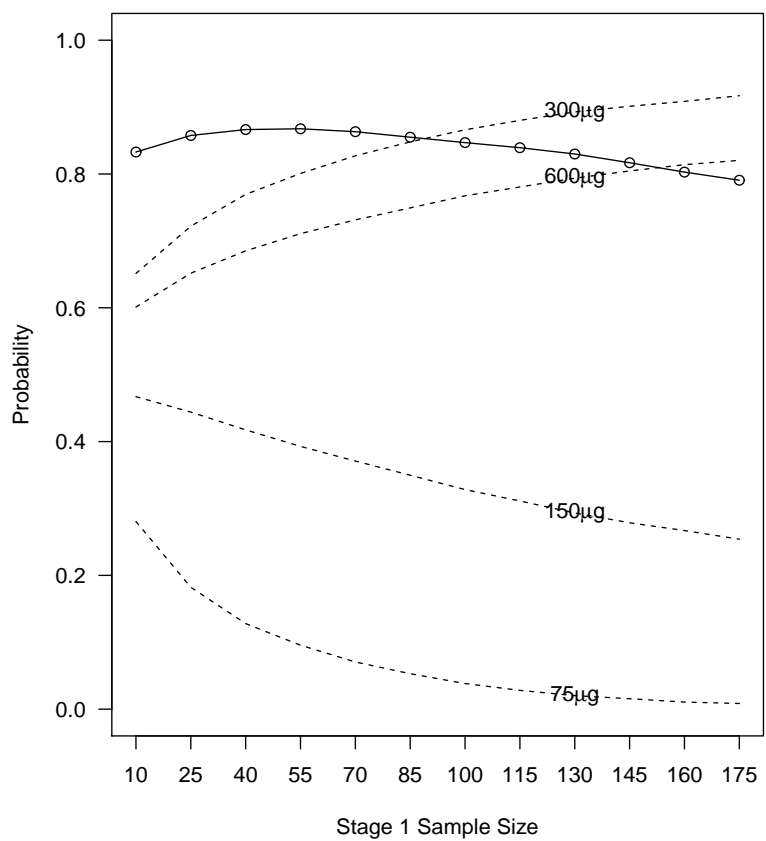}
\caption{Probability of rejection of at least one elementary hypothesis ($\circ$; solid line), and selection probabilities (dashed lines) for stage 1 sample sizes in range 10 to 175 per group.} \label{fig_copd_power}
\end{figure}

The results of the simulations suggest that given the large treatment effects on the early outcome (trough FEV1 at 15 days), and the modest effects on the final outcome (days of poor control (\%) over 26 weeks), the greatest power would have been achieved by using a sample size of around 50 per group in stage 1 (see Figure \ref{fig_copd_power}).

\subsubsection{Continuous early and final outcomes: Selecting varying numbers of doses for the second stage} \label{sec_copd_nn}
As an alternative to always selecting the best two performing treatments at interim, we now consider a threshold rule, where all treatments with test statistics at interim analysis above a fixed threshold are taken into stage 2. If no treatments reach the threshold, the study is stopped for futility. Simulations are implemented using the same effect sizes as in setting 1, a stage 1 sample size of 40 and a stage 2 sample size of 400 in the following code for a threshold of 3:

{
\begin{verbatim}
treatsel.sim(n=list(stage1=40,stage2=400),
      effect=list(early=c(0,0.68,0.82,0.95,0.91), 
      final= c(0,0.13,0.17,0.23,0.20)),
      outcome=list(early="N",final="N"),
      nsim=10000,corr=0.4,seed=145514,select=6,
      thresh=3,level=0.025,ptest=c(3,4))
\end{verbatim} }

The ``{\tt select=6}'' options implements the threshold rule, with the fixed early outcome test statistic threshold set using the ``{\tt thresh}'' option. Results from running this code are as follows (omitting the parts describing the setup):

{\footnotesize \begin{verbatim}
simulation of test statistics: 
expectation early = 3 3.7 4.2 4.1 
expectation final stage 1 = 0.6 0.8 1 0.9 and stage 2 = 1.8 2.4 3.3 2.8 
weights: stage 1 = 0.3 and stage 2 = 0.95 

number of treatments selected at stage 1: 
              n               % 
     1      800            8.00 
     2     1634           16.34 
     3     3098           30.98 
     4     4175           41.75 
 Total     9707           97.07 

treatment selection at stage 1: 
              n               % 
     1     5083           50.83 
     2     7469           74.69 
     3     8914           89.14 
     4     8596           85.96 

hypothesis rejection at study endpoint: 
              n               % 
    H1     2480           24.80 
    H2     4882           48.82 
    H3     7769           77.69 
    H4     6642           66.42 

reject H3 and/or H4 = 8600 :  86%
\end{verbatim} }

\begin{figure}[ht]
\centering
\includegraphics[width=12cm]{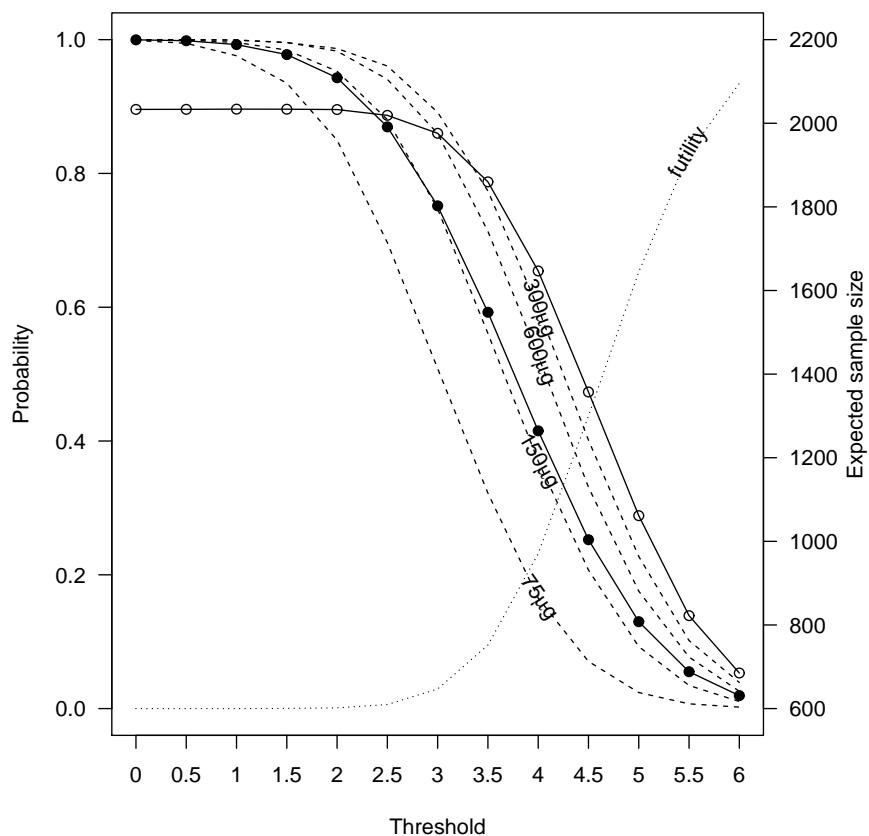}
\caption{Probability of rejection of at least one elementary hypothesis ($\circ$; solid line), futility stopping (dotted line), selection probabilities (dashed lines) and expected total sample size ($\bullet$; solid line) for thresholds on a standardized z-scale in the range 0 to 6.} \label{fig_copd_power_thresh}
\end{figure}

The output shows that this design was stopped for futility only 3\% of the time; with all four experimental treatments being taken into stage 2 more than 41\% of the time. Running the above code for thresholds in the range 0 to 6 (at intervals of a half) and extracting output for each option gives the results summarized in Figure \ref{fig_copd_power_thresh}.

The expected overall sample sizes for the scenarios in Figure \ref{fig_copd_power_thresh}, based on the simulated number of treatments selected at stage 1 and the fixed stage 1 and stage 2 sample sizes of 40 and 400, are as follows; 2199.5, 2197.3, 2188.1, 2164.3, 2109.0, 1991.2, 1802.5, 1548.2, 1264.0, 1004.2, 807.9, 688.2 and 631.0. The power drops of rapidly as the threshold increases from 3 to 5, as futility stopping increases from 3\% to 65\%. When, on average, two test treatments are taken into stage 2, that is when the fixed threshold is somewhere between 3.5 and 4 (overall sample size between 1548.2 and 1264.0), the power is lower (between 78.7\% and 65.4\%) than the analogous setting in Figure 1 (86.6\%). The early outcome effect sizes and distribution are such that a fixed threshold that on average picks two treatments at interim, also stops for futility so often that it reduces the power considerably compared to a design that always takes only two treatments.

\subsubsection{Continuous early and binary final outcome}
The implementation described here has focused on normal outcome measures for both early and final outcomes. However, if one or other outcome were binary the changes necessary to implement this new scenario are relatively straightforward. For instance, instead of using days of poor control over 26 weeks as the final outcome measure, we might use a threshold based on this outcome. If this was the case, then success or failure of the treatment could be determined for each study participant, based on some a priori threshold for the number of days that one might expect to maintain control over the 26 week period. For the sake of example, let us consider the failure rate to be 50\% in the control group, and 45\% ($75 \mu g$), 45\% ($150 \mu g$), 40\% ($300 \mu g$) and 40\% ($600 \mu g$), at each dose respectively. Given that every other facet of the design were the same as the first setting in the COPD example, then this design can be implemented using the following code.

{
\begin{verbatim}
treatsel.sim(n=list(stage1=100,stage2=300),
      effect=list(early=c(0,0.68,0.82,0.95,0.91),
      final=c(0.50,0.45,0.45,0.40,0.40)),
      outcome=list(early="N",final="B"),
      nsim=10000,corr=0.4,seed=145514,select=2,
      level=0.025,ptest=c(3,4))
\end{verbatim} }

This gives an overall rejection probability of 76.99\%, by changing the outcome argument for the {\tt treatsel.sim} function of Section \ref{sec_copd_nn} to {\tt list(early="N",final="B")} and the vector of final effect sizes to {\tt c(0.50,0.45,0.45,0.40,0.40)}.

\subsection{Clinical trials in oncology with subgroup selection} \label{sec_onco}

\citet{Jenkins11} suggested designs for adaptive seamless phase II/III designs for oncology trials using correlated survival endpoints. Designs of this type can be implemented relatively straightforwardly using the subpop.sim function. Here we explore some design properties for a typical scenario from amongst the many that \citet{Jenkins11} explored. We assume early and final time-to-event outcomes with a hazard ratio of 0.6 in the subgroup and 0.9 in the full population for both, and a correlation between endpoints of 0.5; in the setting of an oncology trial, the endpoints might be progression free and overall survival. We set the stage 1 sample size to 100 patients per arm and the stage 2 sample size to 300 patients per arm, if we progress in the full population, and if we progress in the subgroup only to 200 patients per arm. The subgroup prevalence is fixed at 0.3. Using the futility rule for selection at interim, with limits for the subgroup and full population both set to 0, this scenario can be implemented using the following code:

{
\begin{verbatim}
subpop.sim(n=list(stage1=100,enrich=200,stage2=300),
      effect=list(early=c(0.6,0.9),final=c(0.6,0.9)),
      sprev=0.3,outcome=list(early="T",final="T"),
      nsim=10000,corr=0.5,seed=1234,select="futility",
      selim=c(0,0),level=0.025,method="CT-SD")
\end{verbatim} }

The ``{\tt method=CT-SD}'' option implements the combination test method with Spiessens and Debois testing procedure. Given test statistics at interim of $S_1$ and $S_2$ for the subgroup and the full population respectively and rule limits $(l_1,l_2)$, the futility rule implements the following options: (i) if $S_1<l_1$ and $S_2<l_2$, continue with a co-primary analysis, (ii) if $S_1<l_1$ and $S_2>=l_2$, continue in the subgroup only, (iii) if $S_1>=l_1$ and $S_2<l_2$, continue in the full population only and (iv) if $S_1 \ge l_1$ and $S_2 \ge l_2$, stop for futility. Results from running this code are as follows (omitting the parts describing the setup):

{\footnotesize \begin{verbatim}
simulation of test statistics: 
expectation early: sub-pop = -1.46 : full-pop = -0.58 
expectation final stage 1: sub-pop = -1.46 : full-pop = -0.58 
expectation final stage 2: sub-pop only = -3.76 : full-pop only = -1.01 
expectation final stage 2, both groups selected: sub-pop = -2.52 : full-pop = -1.01 
weights: stage 1 = 0.5 and stage 2 = 0.87 

hypotheses rejected and group selection options at stage 1 (n): 
             Hs       Hf      Hs+Hf      Hs+f         n      n% 
sub        2225        0          0        2225    2309   23.09 
full          0       48          0          54     227    2.27 
both       5370     1658       1636        5407    6987   69.87 
total      7595     1706       1636        7686    9523       - 
%         75.95    17.06      16.36       76.86   95.23       - 
reject Hs and/or Hf =  76.65%
\end{verbatim} }

The output reports that in 75.95\%, 17.06\% and 16.36\% of the simulations we rejected in the subgroup, full population and both, respectively. The final two columns give a breakdown of the selections made at the interim analysis; 23.09\% of the simulations were continued in the subgroup only, 2.27\% in the full population only, 69.87\% in both and 4.77\% were stopped for futility. A concise summary of this table can be obtained for further analysis, by assigning to an output object and accessing the results using the syntax {\tt output\$results}.

It is informative in understanding the futility rule, to run the above code for a grid of futility rule limits in the range 0 to -3; the results of this for each option are summarized in Table \ref{tab_onco}, where the notation $l_S$ and $l_F$ indicate the limits for the subgroup and full population, respectively.

\begin{table}[ht]
\centering
\caption{Selection probabilities, probability of futility stopping and power (probability of rejecting at least one elementary null hypothesis) for a range of values of $l_S$ and $I_F$.} \label{tab_onco}
\begin{tabular}{rrrrrrr}
\hline \hline
$l_F$ & $l_S$ & \multicolumn{3}{c}{Selection (\%)} & Futility (\%) & Power (\%) \\
\multicolumn{2}{c}{} & Subgroup & Full & Both & \multicolumn{2}{c}{} \\
\hline 
0	& 0	& 23.1 &	2.3	& 69.9 &	4.8 &	76.7 \\
0	& -1 &	11.4 &	16.2 &	55.8 &	16.7 &	58.8 \\
0 &	-2 &	2.3 &	45.1 &	26.5 &	26.1 &	34.2 \\
0 &	-3 &	0.1 &	66.0 &	6.0 &	27.9 &	20.7 \\
-1 &	0	& 60.0 &	0.4 &	32.3 &	7.3 &	83.9 \\
-1 &	-1 &	37.4 &	4.0 &	29.7 &	29.0 &	61.4 \\
-1 &	-2 &	12.3 &	16.5 &	16.9 &	54.2 &	30.3 \\
-1 &	-3 &	1.5 &	28.4 &	4.8 &	65.4 &	13.8 \\
-2 &	0 &	84.9 &	0.0 &	7.4 &	7.7 &	88.6 \\
-2 &	-1 &	60.1 &	0.3 &	7.2 &	32.4 &	65.0 \\
-2 &	-2 &	24.1 &	2.4 &	5.6 &	68.0 &	29.2 \\
-2 &	-3 &	4.4 &	5.3 &	2.1 &	88.2 &	8.0 \\
-3 &	0	& 91.6 &	0.0	& 0.7 &	7.7 &	89.7 \\
-3 &	-1 &	66.7 &	0.0 &	0.7 &	32.6 &	66.0 \\
-3 &	-2 &	28.6 &	0.1 &	0.6 &	70.7 &	28.8 \\
-3 &	-3 &	5.6 &	0.4 &	0.3 &	93.7 &	6.1 \\
\hline \hline
\end{tabular}
\end{table}

From Table \ref{tab_onco} it is clear that as $l_S$ becomes more negative the subgroup is selected progressively less often and similarly as $l_F$ becomes more negative, then, it is selected progressively more often. The balance between the two limits determines overall power in this setting. With a strong effect in the subgroup and a much weaker effect in the full population, the best strategy is to always, unless stopping for futility, test in the subgroup at the final analysis. For the grid of values tested here this is best achieved when $l_F=-3$ and $l_S=0$.

\section{Discussion}

Adaptive seamless designs are recognized as a tool to increase the efficiency of clinical development programmes by combining features of learning and confirming in a single trial, while traditional development programmes would have investigated these in separate trials. However, their implementation is more involved than traditional designs (see e.g. \citet{Quinlan06} for a discussion). One aspect is the planning which is more complex often requiring extensive Monte Carlo simulations \citep{Benda10, Friede10}. Here we presented a unified framework for adaptive seamless designs with treatment or subpopulation selection. Furthermore, we developed a flexible and yet efficient simulation model. This, as all other methods discussed, can accommodate interim selection informed by an early outcome rather than the final one. Furthermore, we demonstrate how the R package {\tt asd}, freely available from CRAN, can be used to evaluate and compare operating characteristics of various designs.   

There are of course some limitations. Here we focused very much on hypothesis testing, although the estimation of the treatment effects is equally important. There has been some interest in improved estimators in adaptive seamless designs with treatment \citep{Posch05, Brannath09, BowdenGlimm14, StallardKimani18} or subgroup selection \citep{Kimani15, Kimani18}, in particular in more recent years. 

With regard to interim decisions we considered here only fairly straightforward rules although Bayesian statistics (e.g. predictive probabilities \citep{Brannath09}) are also used as the basis for interim decisions. Currently we consider expanding the R package {\tt asd} in this direction.
\vspace*{12pt}

\section*{Acknowledgement}
All authors gratefully acknowledge support by the UK Medical Research Council (grant number G1001344).
\vspace*{1pc}

\noindent {\bf{Conflict of Interest}}

\noindent {\it{The authors have declared no conflict of interest.}}

\section*{Appendix.  Derivation of data simulation models}

We consider first treatment selection designs.  Let $\hat{\mu}^{(i)}_{kj}$ be the estimate for treatment $k$, $k=0, 1, \ldots, K$ with $k=0$ corresponding to control, at stage $j$ , $j=1, \ldots, J$, for endpoint $i$, $i=1,2$.

Assume, as often holds at least asymptotically, that $\hat{\mu}^{(i)}_{kj}$ is normally distributed with 
\[
\hat{\mu}^{(i)}_{kj} \sim N( \mu^{(i)}_k, (\mathscr{I}_{kj}^{(i)})^{-1}),
\]
with
\[
cov(\hat{\mu}^{(i)}_{kj}, \hat{\mu}^{(i)}_{k^\prime j^\prime}) = 0, k \ne k^\prime,
\]
\[
cov(\hat{\mu}^{(i)}_{kj}, \hat{\mu}^{(i)}_{k j^\prime}) = (\mathscr{I}_{k \max\{j,j^\prime\}}^{(i)})^{-1}
\]
and
\[
cov(\hat{\mu}^{(i)}_{kj}, \hat{\mu}^{(i^\prime)}_{k j^\prime}) = \rho \sqrt { (\mathscr{I}_{k \max\{j,j^\prime\}}^{(i)})^{-1} (\mathscr{I}_{k \max\{j,j^\prime\}}^{(i^\prime)})^{-1} }, i \ne i^\prime.
\]

For independent normal random variables with known variance $\sigma^2$, we have $\mathscr{I}_{k \max\{j,j^\prime\}}^{(i)} = \sigma^2/n^{(i)}_{kj}$ where $n^{(i)}_{kj}$ is the number of observations at look $j$ for treatment $k$ on endpoint $i$.

Let $\hat \theta^{(i)}_{kj} = \hat \theta^{(i)}_{kj} - \hat \theta^{(i)}_{0j}, k=1, \ldots, K$, $I^{(i)}_{kj} = ((\mathscr{I}_{kj}^{(i)})^{-1} + (\mathscr{I}_{0j}^{(i)})^{-1}) ^{-1}$ and $S^{(i)}_{kj} = \hat \theta^{(i)}_{kj} I^{(i)}_{kj}$.

Let $\lambda^{(i)}_{kj} = (\mathscr{I}_{kj}^{(i)})^{-1} / (\mathscr{I}_{0j}^{(i)})^{-1}$.  If $\lambda^{(i)}_{kj}$ is constant, say equal to $\lambda$, and $I^{(i)}_{kj} = I^{(i)}_k$ for all $k$ (for a single sample of known-variance normals, this is equivalent to assuming that sample sizes are the same for all experimental treatments).  In this case, considering first a single endpoint, we get

\[
\left (
\begin{array}{c}
S^{(1)}_{11} \\ \vdots \\ S^{(1)}_{K1}\\ \vdots \\ S^{(1)}_{1J} \\ \vdots \\ S^{(1)}_{KJ}
\end{array}
\right )
\sim
N
\left (
\left (
\begin{array}{c}
\theta^{(1)}_1 I^{(1)}_1 \\ \vdots \\ \theta^{(1)}_K I^{(1)}_1\\ \vdots \\ \theta^{(1)}_1 I^{(1)}_J \\ \vdots \\ \theta^{(1)}_K I^{(1)}_J
\end{array}
\right )
,
 \Sigma
\right )
\]

where the variance matrix is given by 
\[
\Sigma = 
\left (
\begin{array}{cccc}
\begin{array}{cccc}
I^{(1)}_1 & \frac{I^{(1)}_1}{1+\lambda} & \cdots & \frac{I^{(1)}_1}{1+\lambda} \\ 
\frac{I^{(1)}_1}{1+\lambda} & I^{(1)}_1 & \ddots & \vdots \\ 
\vdots & \ddots & \ddots & \frac{I^{(1)}_1}{1+\lambda}\\
\frac{I^{(1)}_1}{1+\lambda} & \cdots & \frac{I^{(1)}_1}{1+\lambda} & I^{(1)}_1 \\ 
\end{array}
&
\begin{array}{cccc}
I^{(1)}_1 & \frac{I^{(1)}_1}{1+\lambda} & \cdots & \frac{I^{(1)}_1}{1+\lambda} \\ 
\frac{I^{(1)}_1}{1+\lambda} & I^{(1)}_1 & \ddots & \vdots \\ 
\vdots & \ddots & \ddots & \frac{I^{(1)}_1}{1+\lambda}\\
\frac{I^{(1)}_1}{1+\lambda} & \cdots & \frac{I^{(1)}_1}{1+\lambda} & I^{(1)}_1 \\ 
\end{array}
&
\cdots 
&
\begin{array}{cccc}
I^{(1)}_1 & \frac{I^{(1)}_1}{1+\lambda} & \cdots & \frac{I^{(1)}_1}{1+\lambda} \\ 
\frac{I^{(1)}_1}{1+\lambda} & I^{(1)}_1 & \ddots & \vdots \\ 
\vdots & \ddots & \ddots & \frac{I^{(1)}_1}{1+\lambda}\\
\frac{I^{(1)}_1}{1+\lambda} & \cdots & \frac{I^{(1)}_1}{1+\lambda} & I^{(1)}_1 \\ 
\end{array}
\\ 
\\
\begin{array}{cccc}
I^{(1)}_1 & \frac{I^{(1)}_1}{1+\lambda} & \cdots & \frac{I^{(1)}_1}{1+\lambda} \\ 
\frac{I^{(1)}_1}{1+\lambda} & I^{(1)}_1 & \ddots & \vdots \\ 
\vdots & \ddots & \ddots & \frac{I^{(1)}_1}{1+\lambda}\\
\frac{I^{(1)}_1}{1+\lambda} & \cdots & \frac{I^{(1)}_1}{1+\lambda} & I^{(1)}_1 \\ 
\end{array}
& 
\begin{array}{cccc}
I^{(1)}_2 & \frac{I^{(1)}_2}{1+\lambda} & \cdots & \frac{I^{(1)}_2}{1+\lambda} \\ 
\frac{I^{(1)}_2}{1+\lambda} & I^{(1)}_2 & \ddots & \vdots \\ 
\vdots & \ddots & \ddots & \frac{I^{(1)}_2}{1+\lambda}\\
\frac{I^{(1)}_2}{1+\lambda} & \cdots & \frac{I^{(1)}_2}{1+\lambda} & I^{(1)}_2 \\ 
\end{array}
& 
\cdots 
&
\begin{array}{cccc}
I^{(1)}_2 & \frac{I^{(1)}_2}{1+\lambda} & \cdots & \frac{I^{(1)}_2}{1+\lambda} \\ 
\frac{I^{(1)}_2}{1+\lambda} & I^{(1)}_2 & \ddots & \vdots \\ 
\vdots & \ddots & \ddots & \frac{I^{(1)}_2}{1+\lambda}\\
\frac{I^{(1)}_2}{1+\lambda} & \cdots & \frac{I^{(1)}_2}{1+\lambda} & I^{(1)}_2 \\ 
\end{array}
\\
\\
\vdots & \vdots & \ddots & \vdots \\ \\
\begin{array}{cccc}
I^{(1)}_1 & \frac{I^{(1)}_1}{1+\lambda} & \cdots & \frac{I^{(1)}_1}{1+\lambda} \\ 
\frac{I^{(1)}_1}{1+\lambda} & I^{(1)}_1 & \ddots & \vdots \\ 
\vdots & \ddots & \ddots & \frac{I^{(1)}_1}{1+\lambda}\\
\frac{I^{(1)}_1}{1+\lambda} & \cdots & \frac{I^{(1)}_1}{1+\lambda} & I^{(1)}_1 \\ 
\end{array}
&
\begin{array}{cccc}
I^{(1)}_2 & \frac{I^{(1)}_2}{1+\lambda} & \cdots & \frac{I^{(1)}_2}{1+\lambda} \\ 
\frac{I^{(1)}_2}{1+\lambda} & I^{(1)}_2 & \ddots & \vdots \\ 
\vdots & \ddots & \ddots & \frac{I^{(1)}_2}{1+\lambda}\\
\frac{I^{(1)}_2}{1+\lambda} & \cdots & \frac{I^{(1)}_2}{1+\lambda} & I^{(1)}_2 \\ 
\end{array}
& 
\cdots 
&
\begin{array}{cccc}
I^{(1)}_J & \frac{I^{(1)}_J}{1+\lambda} & \cdots & \frac{I^{(1)}_J}{1+\lambda} \\ 
\frac{I^{(1)}_J}{1+\lambda} & I^{(1)}_J & \ddots & \vdots \\ 
\vdots & \ddots & \ddots & \frac{I^{(1)}_J}{1+\lambda}\\
\frac{I^{(1)}_J}{1+\lambda} & \cdots & \frac{I^{(1)}_J}{1+\lambda} & I^{(1)}_J \\ 
\end{array}
\end{array}
\right )
.
\]

Note that we can rewrite the right hand side of () as
\[
N \left ( I^{(1)} \otimes \theta{(1)}, \Sigma^{(GS)}(I^{(1)}) \otimes \Sigma^{(CS)}_K(\frac{1}{1+\lambda}) \right )
\]
where $I^{(1)}$ and $\theta{(1)}$ denote respectively the vectors $(I^{(1)}_1, \ldots, I^{(1)}_J)^\prime$, and $(\theta{(1)}_1, \ldots, \theta{(1)}_K)^\prime$ and 
\[
\Sigma^{(CS)}_K(r) = 
\left (
\begin{array}{cccc}
1 & r & \cdots & r \\ 
r & 1 & \ddots & \vdots \\ 
\vdots & \ddots & \ddots & r\\
r & \cdots & r & 1 \\ 
\end{array}
\right )
\]
and 
\[
\Sigma^{(GS)}(I)^{(1)} = 
\left (
\begin{array}{cccc}
I^{(1)}_1 & I^{(1)}_1 & \cdots & I^{(1)}_1 \\ 
I^{(1)}_1 & I^{(1)}_2 & \cdots & I^{(1)}_2 \\
\vdots & \vdots & \ddots & \vdots \\
I^{(1)}_1 & I^{(1)}_2 & \cdots & I^{(1)}_J\\
\end{array}
\right )
\]
 denote variance matrices of the $K \times K$ complex symmetric form and of that obtained in the usual group-sequential setting.

When both endpoints are considered, the vector 
\[
(S^{(1)}_{11}, \ldots, S^{(1)}_{K1}, \ldots, S^{(1)}_{1J}, \ldots, S^{(1)}_{KJ},   S^{(2)}_{11}, \ldots, S^{(2)}_{K1}, \ldots, S^{(2)}_{1J}, \ldots, S^{(2)}_{KJ})^\prime
\]
 is normally distributed with mean 
\[
\left (
\begin{array}{c}
\theta^{(1)} \otimes I^{(1)} \\
\theta^{(2)} \otimes I^{(2)} 
\end{array}
\right )
\]
and variance
\[
\left (
\begin{array}{cc}
\Sigma^{(GS)}(I^{(1)}) & \rho \Sigma^{(GS)}(I^{(12)}) \\
\rho \Sigma^{(GS)}(I^{(12)}) & \Sigma^{(GS)}(I^{(2)})
\end{array}
\right )
\otimes \Sigma^{(CS)}_K(\frac{1}{1+\lambda})
\]
where is a $J$-dimensional vector with $I^{(12)}_j = \sqrt{I^{(1)}_j I^{(2)}_j}, j=1, \ldots, J$.

We consider next designs with subgroup selection.  In analogy to the notation used above, now let $\hat{\theta}^{(i)}_{kj}$ be the estimate the treatment effect in subgroup $k$, $k=1, \ldots, K$ at stage $j$ , $j=1, \ldots, J$, for endpoint $i$, $i=1,2$.  We will typically consider the case of $K=2$, with $k=2$ corresponding to the entire population and $k=1$ corresponding to a subgroup comprising a proportion $\tau$ of the entire population.  

Let $(I^{(i)}_{kj})^{-1}$ denote the variance of $\hat{\theta}^{(i)}_{kj}$ and let $S^{(i)}_kj = \hat{\theta}^{(i)}_{kj} I^{(i)}_{kj}$.  

As above, we will assume that the $\hat{\theta}^{(i)}_{kj}$ are normally distributed with mean $\theta^{(i)}_k$ and variance $(I^{(i)}_{kj})^{-1}$ with
\[
cov ( \hat{\theta}^{(i)}_{kj}, \hat{\theta}^{(i)}_{k^\prime j^\prime} )= (I^{(i)}_{\max\{k, k^\prime\} \max\{j, j^\prime\}})^{-1}
\]
and
\[
cov ( \hat{\theta}^{(i^\prime)}_{kj}, \hat{\theta}^{(i)}_{k^\prime j^\prime} ) = \rho (I^{(i)}_{\max\{k, k^\prime\} \max\{j, j^\prime\}} I^{(i)}_{\max\{k, k^\prime\} \max\{j, j^\prime\}})^{-1/2}, i\ne i^\prime.
\]

Denote $I^{(i)}_{j2}$ by $I^{(i)}_j$ and assume further that $I^{(i)}_{j1} = \tau I^{(i)}_j$ for $i=1,2, j=1, \ldots, J$ (see Spiessens and Debois), then 
\[
(S^{(1)}_{11}, S^{(21)}_{1}, \ldots, S^{(1)}_{1J}, S^{(1)}_{2J},   S^{(2)}_{11}, S^{(2)}_{21}, \ldots, S^{(2)}_{1J}, S^{(2)}_{2J})^\prime
\]
 is normally distributed with mean 
\[
\left (
\begin{array}{c}
\theta^{(1)} \otimes I^{(1)} \\
\theta^{(2)} \otimes I^{(2)} 
\end{array}
\right )
\]
and variance
\[
\left (
\begin{array}{cc}
\Sigma^{(GS)}(I^{(1)}) & \rho \Sigma^{(GS)}(I^{(12)}) \\
\rho \Sigma^{(GS)}(I^{(12)}) & \Sigma^{(GS)}(I^{(2)})
\end{array}
\right )
\otimes \Sigma^{(GS)}\left ( \begin{array}{c} \tau \\ 1 \end{array} \right )
\]
where $I^{(i)} = (I^{(i)}_1, \ldots, I^{(i)}_J)^\prime$, $I^{12} = (I^{(12)}_1, \ldots, I^{(12)}_J)^\prime$ with $I^{(12)}_j = \sqrt{I^{(1)}_j I^{(2)}_j}$, $\theta^{(i)} = (\theta^{(i)}_1, \theta^{(i)}_2)$ and $\Sigma^{(GS)}(I)$ and $\Sigma^{(CS)}_K(r)$ are as defined above.

If more than two subgroups are considered, or if these are nested in some other way than the second being a subset of the first, the last matrix can be modified accordingly to reflect this.

\end{document}